# Van der Waals heterostructures for spintronics and opto-spintronics

Juan F. Sierra[1*], Jaroslav Fabian[2], Roland K. Kawakami[3], Stephan Roche[1,4] and Sergio O. Valenzuela[1,4*]

[1]*Catalan Institute of Nanoscience and Nanotechnology (ICN2), CSIC and The Barcelona Institute of Science and Technology (BIST), Campus UAB, Bellaterra, 08193 Barcelona, Spain*

[2]*Institute for Theoretical Physics, University of Regensburg, 93040 Regensburg, Germany*

[3]*Department of Physics, The Ohio State University, Columbus, OH 43210, USA*

[4]*Institució Catalana de Recerca i Estudis Avançats (ICREA), 08010 Barcelona, Spain*

*e:mail: juan.sierra@icn2.cat; SOV@icrea.cat

**The large variety of 2D materials and their co-integration in van der Waals (vdW) heterostructures enable innovative device engineering. In addition, their atomically-thin nature promotes the design of artificial materials by proximity effects that originate from short-range interactions. Such a designer approach is particularly compelling for spintronics, which typically harnesses functionalities from thin layers of magnetic and non-magnetic materials and the interfaces between them. Here, we overview recent progress on 2D spintronics and opto-spintronics using vdW heterostructures. After an introduction to the forefront on spin transport research, we highlight the unique spin-related phenomena arising from spin-orbit and magnetic proximity effects. We further describe the ability to create multi-functional hybrid heterostructures based on vdW materials, combining spin, valley and excitonic degrees of freedom. We end with an outlook on perspectives and challenges for the design and production of ultra-compact all-2D spin devices and their potential applications in conventional and quantum technologies.**

In the past few years, van der Waals (vdW) heterostructures[1,2] comprising a variety of 2D layered materials have emerged as potential building blocks for future ultrafast and low-power electronic and spintronic devices. Graphene is an ideal spin channel owing to its spin diffusion length reaching several micrometres at room temperature, gate-tunable carrier concentration and extremely high carrier mobility[3–5]. Semiconducting TMDCs[6] such as $MX_2$ (M= W, Mo; X = S, Se, Te) and topological insulators (TIs) such as $Bi_2Te_3$ possess strong spin-orbit coupling (SOC), which allows for the electrical generation and manipulation of spins. Semiconducting TMDCs further possess a strong spin-photon coupling that enables optical spin injection, while 2D magnets[7] bring capabilities for spin filtering and non-volatile data storage.

Novel functionalities arise due to the atomically-thin nature of 2D materials, which facilitates much stronger electrostatic gating effects than with conventional materials to achieve, for instance, voltage-controlled magnetism. Furthermore, the integration of graphene, TMDC, TIs and 2D magnets into vdW heterostructures not only combines the respective material functionalities but also imprints properties through proximity interactions across interfaces[8], enabling the design of artificial structures with unique characteristics. Such properties provide opportunities[9,10] for memory applications, spin interconnects, spin-transistors, microwave nano-oscillators, low-power reconfigurable logics and for flexible or wearable spintronic platforms[11,12] (see Box 1).

This Review presents the state of the art and future prospects for vdW heterostructures in spintronics and optospintronics, with a special focus on magnetic and spin-orbit proximity effects and the emerging phenomena deriving from them. Covering recent experimental and theoretical developments, the Review is divided in four main sections. The first section briefly surveys the recent progress in spin injection and detection, including the integration of opto-electronic elements, and then outlines the contemporary understanding of spin dynamics in 2D materials. This description is complemented by an overview of materials that can be utilized to enhance the spin properties or further create multifunctional 2D spintronic devices (Box 1). This is followed by the core of the Review on vdW heterostructures and proximity effects. First, the second section focuses on proximity-induced SOC, which is central in modern spintronics as it can enhance the magnetic properties of 2D magnets as well as provide spin filtering, spin manipulation and efficient charge to spin interconversion (CSI) functionalities. Next, the third section addresses magnetic proximity effects, which can be harnessed in memory elements, reconfigurable spin-logic circuits and novel spin-valleytronics applications. Besides the vast catalogue of material combinations, vdW heterostructures establish new concepts based on twist-angle and stacking control between crystallographic lattices that can strongly dictate the nature and strength of proximity phenomena. Finally, the fourth section discusses potential applications and future research directions and perspectives.

## Spin dynamics in 2D materials

**Recent advances on spin injection and detection.** Spin dynamics is typically investigated using lateral devices in a non-local electrical configuration[3,5] or, alternatively, using spectroscopic methods in optically active materials[13]. Lateral spin devices rely on efficient spin injection and detection with tunnel barriers playing a crucial role in alleviating the conductance mismatch problem[14], which limits the effective spin polarization $P_s$ of the injector and detector contacts (Fig. 1a). Early studies with graphene as a spin channel used MgO, $Al_2O_3$, $TiO_2$, and amorphous carbon barriers[15–17] but the emergence of alternative insulators could improve the device performance in various aspects. SrO barriers, grown by evaporation of Sr in the presence of molecular oxygen, lead to robust operation with high bias (~2 V) to achieve large spin accumulation[18]. Barriers composed of 2D materials produce high $P_s$. Fluorinated graphene, obtained by exposure to $XeF_2$ gas, yields $P_s > 40\%$[19] while hexagonal boron nitride (hBN), using stacking and transfer methods, efficiently injects spins into graphene[20,21] and black phosphorous[22]. With graphene, hBN-based injectors display a *differential* spin polarization that varies with applied voltage bias, reaching absolute values above 100 % and even changing sign[21]. Beyond planar contacts, one-dimensional edge contacts have been utilized for spin injection into hBN-encapsulated graphene[23].

**Box 1| Designer van der Waals (vdW) heterostructures for spintronics. Material functionalities.**

Key practical elements in spintronics are the injection, transport or communication, manipulation, and detection of spin information[10,13,24,25]. Each of these aspects, represented in the figure below, can benefit from the unique properties of 2D materials. While spin communication requires low spin-orbit materials, spin injection, detection and manipulation can be achieved by creating multifunctional vdW heterostructures, which include materials with magnetic exchange interaction (MEI) and/or large spin-orbit coupling (SOC).

**Spin communication** over practical distances has already been demonstrated in graphene[26,27] and black phosphorous[22] at room temperature, being enhanced by vdW encapsulation with hexagonal boron nitride (hBN)[28]. Spin logics and multiplexer devices have been proposed, using spin accumulation[29], gate dependence of spin lifetimes and/or drift currents[30,31].

**Spin injection and detection** use several approaches[10,24]. They involve insulating 2D ferromagnets (2DFMs), metallic 2DFMs in combination with insulating tunnel barriers, or topological insulators (TIs) and other large-SOC materials through charge-to-spin interconversion (CSI) mechanisms, such as the spin Hall and spin galvanic effects (SHE, SGE).

Optical spin injection in graphene, using transition metal dichalcogenides[32] (TMDCs), has also been demonstrated[33,34]. MTJs, combining 2DFMs and insulating hBN barriers, or tunnel filters comprising insulating (anti)ferromagnetic materials, could be used in ultra-compact low-power memory elements[10,24].

**Spin manipulation and control** as well as spin injection and detection can be engineered via **proximity-induced SOC and proximity-induced MEI** by adjacent (insulating) large-SOC materials (*e.g*, TMDCs and TIs) and (anti)FMs, respectively. Magnetization switching and precession in memory elements could be achieved with SOTs[10], using CSI in materials with large-SOC, either of intrinsic origin or acquired by proximity effects. Novel spin-transistor configurations and spin polarizers, which take advantage of proximitized spin-valley coupling in graphene, have been experimentally demonstrated[35–37]. The atomically thin nature of 2D (anti)FMs further enable the manipulation of their magnetic state using electric fields[38], while twist-angle and stacking control between crystallographic lattices add yet other versatile knobs to engineer the nature and strength of proximity effects.

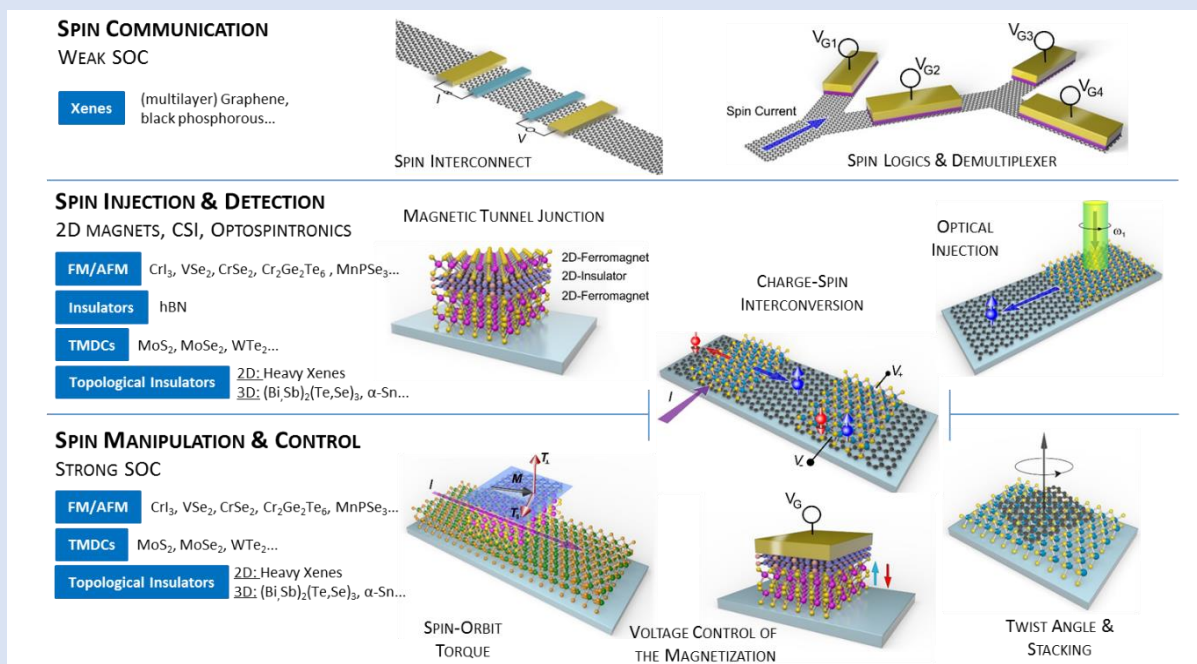

Optical selection rules together with Faraday and Kerr microscopies, broadly used in semiconducting materials[13], are valuable tools to investigate spin-valley dynamics and spin coherence in TMDCs[39–41]. Typically, a pump laser illuminates individual crystals with right- or left-circularly polarized light at specific wavelengths to target an exciton transition, generating spin-polarized electrons and holes in the *K* or *K'* valley, respectively (Fig. 1b). The spin polarization can be then detected by means of the optical Kerr rotation of a linearly polarized laser probe.

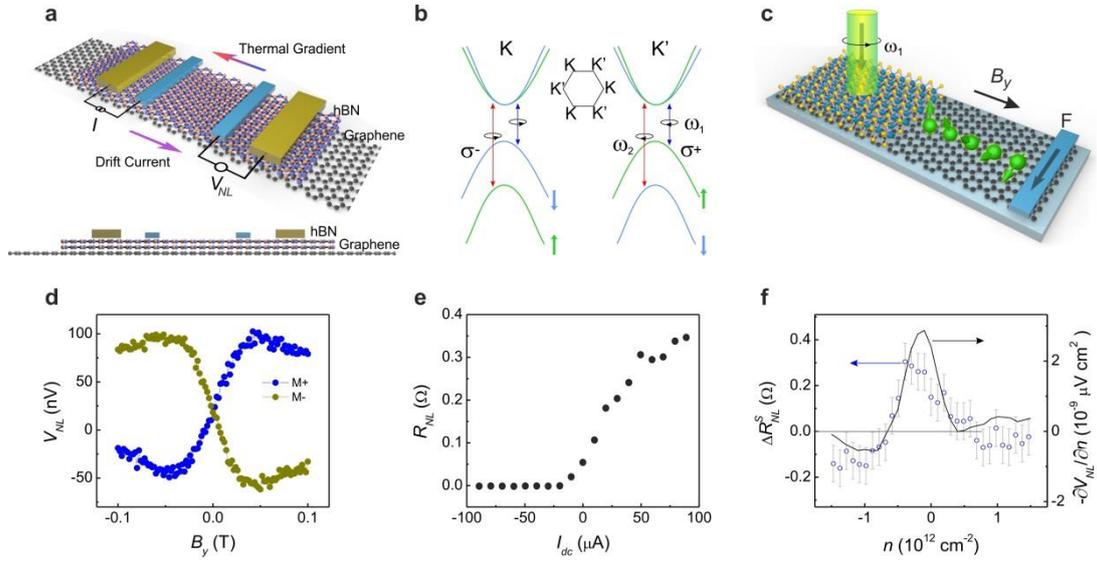

**Fig. 1 | Spin injection and spin transport. a,** Illustration of a lateral graphene-hBN spin device with current and voltage leads in the non-local configuration scheme and control by electric and thermal drifts. Blue (inner) contacts are ferromagnetic, while yellow (outer) contacts are preferably non-magnetic. **b,** Schematic band structure and optically-excited transitions of a monolayer TMDC at the *K* and *K'* valleys with left ($\sigma^-$) and right ($\sigma^+$) circularly polarized light. **c,** Optical spin injection from monolayer TMDC into graphene and subsequent lateral spin transport. The spin current is remotely detected using the spin-sensitive (ferromagnetic) contact F. The injected spins are initially oriented out-of-plane and rotate under the influence of an applied magnetic field $B_y$ as they diffuse towards F, eventually becoming aligned with its magnetization. **d,** Non-local voltage $V_{NL}$ measured at probe F in **c** as a function of $B_y$; $V_{NL}$ exhibits an antisymmetric spin precession signal whose sign changes by flipping the magnetization ***M*** of the detector F. **e,** Spin signal $R_{NL}$= $V_{NL}/I$ as a function of the drift current $I_{dc}$ applied along the spin channel as shown in **a**. The modulation is driven by a change in the effective $\lambda_s$, which increases or decreases when the current flows in favour of or against the spin current, respectively. **f,** Spin signal change $\Delta R_{NL}^S$ as a function of the graphene carrier density *n* induced by the presence of a thermal gradient in the spin channel as shown in **a**. Panel **d** and **e** adapted from Refs. [33] and [42] respectively, American Chemical Society. Panel **f** reproduced from Ref [43], SNL.

When embedded in vdW heterostructures, TMDCs enable a new platform for optospintronics[32] (Box 1). As demonstrated in MoS$_2$- and WSe$_2$-graphene heterostructures, the spin-polarized carriers generated in the TMDC transfer into the neighbouring graphene[33,34] (Fig. 1c). The resulting antisymmetric Hanle spin precession curve (Fig. 1d) under the influence of an applied magnetic field $B_y$ provides unambiguous proof of the optical spin injection.

**Spin dynamics and relaxation.** The spin propagation is characterized by the spin relaxation length $\lambda_s$, given by $\lambda_s = \sqrt{D_s \tau_s}$ with $D_s$ the spin diffusion constant, and $\tau_s$ the spin lifetime. In the diffusive regime, $D_s$ is obtained from transport measurements, while various possible SOC mechanisms, either intrinsic or extrinsic, introduce sources of spin relaxation and dictate the ultimate value of $\tau_s$ and $\lambda_s$[13]. At room temperature, graphene displays spin-transport figures of merit for spin communication that outperform those of all other materials (Box 1). In hBN-protected graphene $\tau_s$ can be larger than 10 ns (with $\lambda_s \sim 30$ μm)[28]. Room-temperature spin-diffusion lengths reaching 10 μm were further achieved in chemical vapour deposition (CVD)-grown graphene on silicon oxide ($SiO_2$) substrates[27]. Black phosphorous also transports spins quite efficiently; when encapsulated with hBN, $\tau_s \sim 0.7$ ns and $\lambda_s \sim 2.5$ μm at room temperature[22]. In TMDCs, spin relaxation has been investigated using optical orientation and time-resolved Kerr rotation. Reported spin-valley lifetimes at low temperature (a few kelvin) exceed several nanoseconds in electron-doped CVD-grown $MoS_2$ and $WS_2$ monolayers[39,44] and are about 80 ns for holes in MOCVD $WSe_2$ monolayers[45]. Although they are significantly longer in exfoliated $WSe_2$ (100 ns for electrons[40] and 1 μs for holes[40,41]) all-temperature dependent studies show a fast decrease with temperature. A strikingly different behaviour has been observed in $MoSe_2$, with the largest lifetime ~100 ns found at room temperature, albeit likely corresponding to non-itinerant carriers[46]. Efficient generation of pure and locked spin-valley diffusion current was demonstrated in exfoliated $WS_2$-$WSe_2$ heterostructures at 10 K by pump-probe spectroscopy[47]. Excitons are created in $WSe_2$ and subsequent fast transfer of excited electrons to $WS_2$ suppresses the exciton-valley depolarization channel. The recombination of electrons in $WS_2$ with holes in $WSe_2$ leaves an excess of holes in one of the $WSe_2$ valleys, which are found to live for more than 20 μs and propagate over 20 μm.

The mechanisms leading to spin relaxation in 2D materials are very rich and frequently unique to each material. This is illustrated by the case of graphene. Theoretical calculations describe a wide range of possible SOC sources, through the symmetry, spatial range and strength of spin-conserving and non-spin conserving events. Intrinsic and Rashba contributions give rise to a small spin-splitting[48] of tens of μeV, as corroborated experimentally[49]. Early theoretical work indicated that $\tau_s$ could be in the millisecond range[3]. However, follow-up studies, introducing realistic descriptions of impurities (magnetic defects such as hydrogen adsorbents serving as spin-flip resonant scatterers[50,51]) or subtle mechanisms such as spin-pseudospin coupling[52,53], have provided alternative explanations for the observed $\tau_s$, in the ns and sub-ns range. These studies account for the $\tau_s$ energy dependence, with the most universal feature being a minimum near the charge neutrality point. The underlying origin for the spin relaxation has been described using the Elliot-Yafet[54] or Dyakonov-Perel[55] mechanisms, however only strictly applicable in disordered systems with short mean free paths[56]. Some progress in analysing spin dynamics in the ballistic limit, as well as possible fingerprints in spin precession measurements, has been made[57]. In polycrystalline graphene, theoretical analysis has revealed universal spin diffusion lengths dictated by the absolute strength of the substrate-induced Rashba SOC in the Dyakonov-Perel regime[56] ($\lambda_s = \hbar v_\mathrm{F}/2\lambda_\mathrm{R}$ with $\hbar$ the Planck constant, $v_\mathrm{F}$ the Fermi velocity, and $\lambda_\mathrm{R}$ the Rashba SOC strength). Despite important progress, a full correspondence between theory and experiment is still missing. Indeed, the predominance of Rashba SOC in spin transport should manifest in a spin-transport anisotropy[56], where the out-of-plane spin lifetime $\tau_{s,\perp}$ is half the in-plane one $\tau_{s,\parallel}$. An electric-field modulation of the spin relaxation anisotropy ratio $\zeta = \frac{\tau_{s,\perp}}{\tau_{s,\parallel}}$, consistent with the presence of Rashba SOC, was reported in graphene encapsulated with hBN[58]. However, it has been argued[59,60] that the application of large out-of-plane magnetic fields could affect the determination of $\zeta$. Recent experimental studies[59–62] have failed to establish a significant spin lifetime anisotropy, suggesting that either magnetic resonant spin-flip scattering or deformation-induced gauge pseudo-magnetic fields randomize the spatial direction of the effective SOC field[50,51,59,61]. Remarkably, as discussed below, a known SOC can be made dominant in proximitized graphene and, in contrast to graphene, the spin dynamics be well understood.

Beyond graphene, the progress has been modest. Spin relaxation in few-layer black phosphorous appears to follow the Elliot-Yafet mechanism, as suggested by the similar temperature dependence of the measured $\tau_s$ and the momentum lifetime[22]. In TMDCs, the long spin-valley lifetimes confirm the expectation of spin-valley locking, which manifests more strongly in the valence band. The relaxation is expected to be mediated by intravalley decoherence mechanisms, dominating electron spin lifetime, and spin-flip processes between valleys, requiring simultaneous scattering of both valley and spin degrees of freedom and yielding slow relaxation rates for holes. As temperature increases the behaviour becomes increasingly complex as relaxation pathways involving secondary valleys and different phonon-mediated intervalley scattering rates may play a role in determining the spin lifetimes[46].

**Current and thermal spin current drift.** Large $\lambda_s$ may facilitate the realization of all-spin reprogrammable operations by controlling spin currents in lateral devices[3,63] (Box 1). In this regard, a XOR magnetologic gate has been experimentally demonstrated at room temperature by electrical bias tuning of the spin injection in graphene[29]. This demonstration was followed by the proposal of a gate-driven demultiplexer using local voltage gates to tune the spin currents[30]. Further experimental progress has been achieved in the control of spin currents via carrier and thermal drift effects. Lateral drift fields in bilayer graphene (BLG), caused by a charge current (Fig. 1a), were shown to modulate the spin signal at room temperature[42] (Fig. 1e). More recently, the use of thermal gradients to enhance or suppress the spin signal has been proposed and demonstrated[43] (Fig. 1f). Here the spin signal modulation is driven by thermal drifts (Fig. 1a) in combination with an energy-dependent thermoelectric power, which result in a thermoelectric spin voltage. The observation of this phenomenon requires sufficiently large lateral thermal gradients, which can be achieved by hot carrier generation, either by electrical current flow in graphene or through tunnel barrier injection[43,64].

**Spin-orbit proximity effects**

Proximity effects represent a versatile approach to material design that can reach its full potential with vdW heterostructures, where hybridization of electronic orbitals of adjacent atomically-thin layers occurs. Despite the weak nature of vdW interaction, interlayer coupling of pure tunnelling character can drastically change the energy dispersion and spin texture of the electronic band structure. For instance, in BLG such tunnelling turns the linear dispersion of low energy excitation to a parabolic shape, in addition to other band modifications. In a trilayer structure, as that represented in Fig. 2a, the intercalated material acquires properties from the top and bottom layers, bringing unprecedented opportunities for spintronics, particularly to imprint a SOC or magnetic exchange interaction (MEI). SOC is ubiquitous in spintronics[10,13] (Box 1), playing a central role in spin relaxation and manipulation, CSI, anisotropic magnetoresistance, perpendicular magnetic anisotropy, spin-orbit torques (SOT) and the emergence of topological states. Therefore proximity SOC concepts are particularly relevant as they can potentially help engineer and control many of these phenomena.

Graphene and BLG represent model materials for proximity-effect studies. In their isolated states, the SOC strength is only tens of μeV and opens a very small spin-orbit gap, as shown by *ab-initio* calculations[48,65,66] and recent experiments[49,67]. The Hamiltonian of isolated graphene is $H \sim H_0$, with $H_0$ characterizing Dirac carriers (see Box 2). When graphene gets in contact with other materials---of interest are 2D semiconductors and 2D insulators that preserve the Dirac cones in their band gaps---the character of $H$ can radically change. Surprisingly generic Hamiltonian models $H = H_0 + H_{t,b}$ have been derived, which capture first-principles results, with $H_{t,b}$ comprising separate orbital, spin-orbit and exchange terms (Box 2) that can be tracked back to the top (*t*) and/or bottom (*b*) 2D materials (Fig. 2a).

**Box 2 | Emergent Hamiltonian in proximitized graphene.**

The emergent Hamiltonian $H$ describing Dirac electrons in proximitized graphene comprises orbital, spin-orbit, and exchange terms:

$$H = H_0 + H_\Delta + H_I + H_{VZ} + H_R + H_{PIA} + H_{EX}$$

$H_0 = \hbar v_F(\tau k_x \sigma_x + k_y \sigma_y)$ is the low-energy graphene Hamiltonian, with $v_F$ Fermi velocity, $k$ the state wave vector and $\sigma$ the pseudospin Pauli matrices. The factor $\tau$ is 1 (-1) at $K$ ($K'$) valleys. $H_\Delta = \Delta \sigma_z$ describes orbital gap opening when the pseudospin symmetry is broken. Other terms include the intrinsic SOC, existing already in pristine graphene, $H_I = \lambda_I \tau \sigma_z s_z$, parametrized by $\lambda_I$, with $s$ the spin Pauli matrices. The valley-Zeeman SOC $H_{VZ} = \lambda_{VZ} \tau \sigma_0 s_z$ with strength $\lambda_{VZ}$ emerges, e.g., when graphene is interfaced with TMDCs or TIs. The Rashba SOC $H_R = \lambda_R(\tau \sigma_x s_y - \sigma_y s_x)$ with strength $\lambda_R$ is present whenever space-inversion symmetry breaks, owing to an electric field or in heterostructures.

The pseudospin inversion asymmetry (PIA) SOC $H_{PIA} = a(\lambda_{PIA}\sigma_z + \Delta_{PIA})(k_x s_y - k_y s_x)$, with $a$ the graphene lattice constant. The terms proportional to $\lambda_{PIA}$ and $\Delta_{PIA}$ lead to a renormalization of the Fermi velocity and a $k$-linear band splitting, respectively. Finally, the proximity exchange coupling $H_{EX} = \lambda_{EX} s_z + \lambda_{EX}^{AF} \sigma_z s_z$ is parametrized by $\lambda_{EX}, (\lambda_{EX}^{AF})$ and emerges when graphene forms heterostructures with ferromagnets (antiferromagnets). Neglecting many-body effects, essentially the same Hamiltonian ($H$) describes the electronic states in (proximitized) TMDCs but instead of sublattice degrees of freedom, $\sigma$ matrices describe the valence and conduction bands[68,69]. The table presents representative graphene-based heterostructures, their electronic band structure at the $K$ (and $K'$) points and the most relevant parameters for each of them by fitting the ab-initio results of the relaxed structures (see corresponding references for further details).

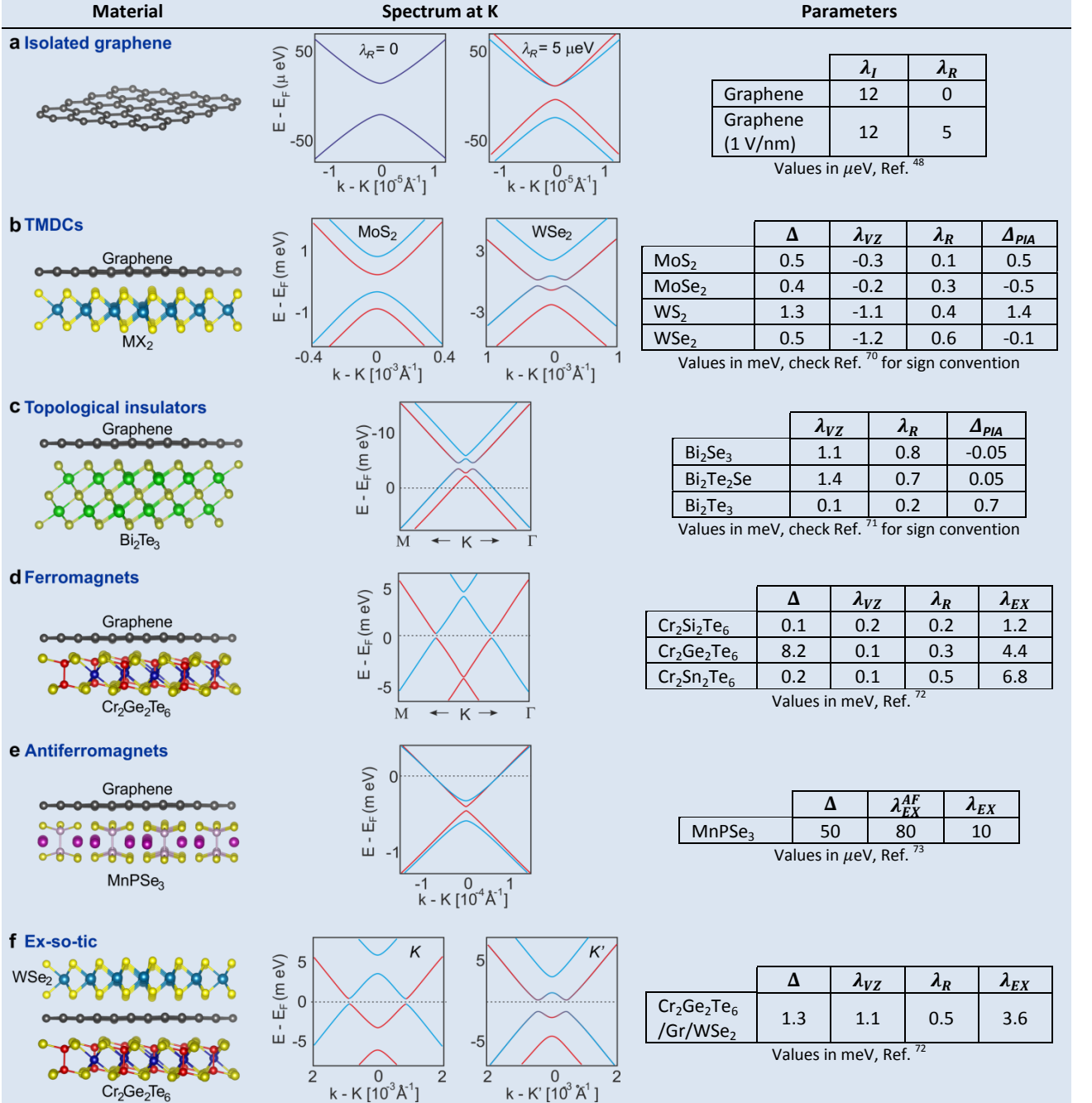

| Material | Spectrum at K | Parameters |
|---|---|---|

**a Isolated graphene**

| | $\lambda_I$ | $\lambda_R$ |
|---|---|---|
| Graphene | 12 | 0 |
| Graphene (1 V/nm) | 12 | 5 |

Values in $\mu$eV, Ref. [48]

**b TMDCs**

| | $\Delta$ | $\lambda_{VZ}$ | $\lambda_R$ | $\Delta_{PIA}$ |
|---|---|---|---|---|
| MoS$_2$ | 0.5 | -0.3 | 0.1 | 0.5 |
| MoSe$_2$ | 0.4 | -0.2 | 0.3 | -0.5 |
| WS$_2$ | 1.3 | -1.1 | 0.4 | 1.4 |
| WSe$_2$ | 0.5 | -1.2 | 0.6 | -0.1 |

Values in meV, check Ref. [70] for sign convention

**c Topological insulators**

| | $\lambda_{VZ}$ | $\lambda_R$ | $\Delta_{PIA}$ |
|---|---|---|---|
| Bi$_2$Se$_3$ | 1.1 | 0.8 | -0.05 |
| Bi$_2$Te$_2$Se | 1.4 | 0.7 | 0.05 |
| Bi$_2$Te$_3$ | 0.1 | 0.2 | 0.7 |

Values in meV, check Ref. [71] for sign convention

**d Ferromagnets**

| | $\Delta$ | $\lambda_{VZ}$ | $\lambda_R$ | $\lambda_{EX}$ |
|---|---|---|---|---|
| Cr$_2$Si$_2$Te$_6$ | 0.1 | 0.2 | 0.2 | 1.2 |
| Cr$_2$Ge$_2$Te$_6$ | 8.2 | 0.1 | 0.3 | 4.4 |
| Cr$_2$Sn$_2$Te$_6$ | 0.2 | 0.1 | 0.5 | 6.8 |

Values in meV, Ref. [72]

**e Antiferromagnets**

| | $\Delta$ | $\lambda_{EX}^{AF}$ | $\lambda_{EX}$ |
|---|---|---|---|
| MnPSe$_3$ | 50 | 80 | 10 |

Values in $\mu$eV, Ref. [73]

**f Ex-so-tic**

| | $\Delta$ | $\lambda_{VZ}$ | $\lambda_R$ | $\lambda_{EX}$ |
|---|---|---|---|---|
| Cr$_2$Ge$_2$Te$_6$/Gr/WSe$_2$ | 1.3 | 1.1 | 0.5 | 3.6 |

Values in meV, Ref. [72]

Owing to the short range of the magnetic-exchange and spin-orbit interactions, the proximity effects are largely driven by the layer adjacent to the proximitized graphene. Therefore, the thickness of the 2D magnet or large-SOC material does not require control. In addition, the proximity effect in BLG predominantly develops only in the layer in contact with the material.

According to the previous discussion, it is not surprising that proximity SOC concepts are best established for graphene. While materials such as hBN do not increase graphene's SOC beyond tens-of-μeV[74], it has been demonstrated that strong SOC materials like as TMDCs or the $Bi_2Se_3$ TI family, significantly alter it or reinforce it (Box 2 b and c). The graphene Dirac cones are preserved within the band-gaps of many TMDCs[70], which allows one to exploit the advantages of graphene's high-mobility and novel proximity spin interactions[75,76]. The SOC strength can reach meVs (Box 2 b) and be dominated by a valley-Zeeman SOC, which is characterized by an out-of-plane spin-orbit field that is opposite at *K* and *K'* valleys (as in the TMDC). In addition, carriers experience a Rashba SOC, with an in-plane spin-orbit field texture perpendicular to the momentum. On the orbital level, the breaking of the pseudospin symmetry leads to the appearance of an orbital gap, described by a staggered potential. The valley-Zeeman and Rashba fields are predicted to change by twisting the graphene relative to the TMDC, with the largest SOC strength appearing at 15-20 degrees between the lattice vectors[77,78]. Band structures at smaller twist angles were also theoretically investigated[79,80]. Graphene can also be proximitized by TIs, such as $Bi_2Se_3$[71,81,82]. These 3D TIs exhibit protected surface states with in-plane spin-orbit fields inducing spin-momentum locking. Surprisingly, the proximitized-SOC is still dominated by the out-of-plane valley-Zeeman coupling, which is not present in the TI (Box 2 c).

Experimental signatures of proximity-induced SOC in graphene-TMDC heterostructures have been found in weak (anti-)localization measurements[83–87]. However, the results are controversial in terms of the SOC strength, which ranges from ~1 to 10 meV, as well as the SOC nature, which was reported to have Rashba or valley-Zeeman character. Variations on the SOC strength may be due in part to variations in the interface properties; twist angle or the presence of trapped bubbles – all of them difficult to control during device fabrication. Nonetheless, the nature of the proximity SOC has been established by means of spin relaxation anisotropy and CSI experiments, as discussed below.

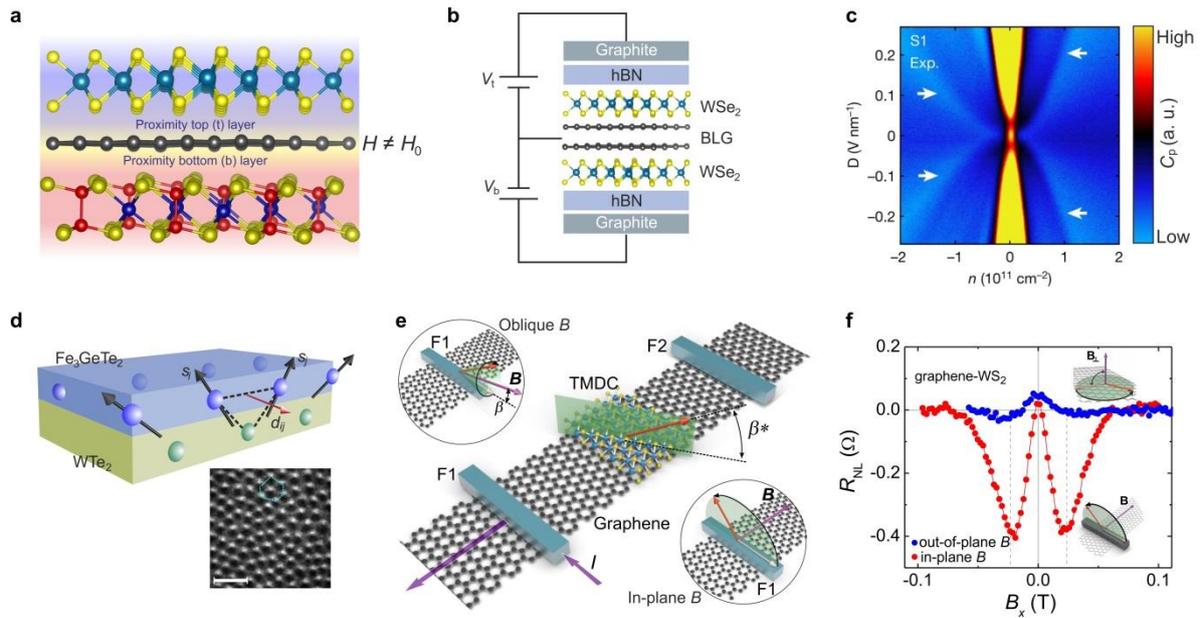

**Fig. 2 | Proximity-effects. a**, In a trilayer vdW heterostructure, the middle layer is proximitized by the top ($t$) and bottom ($b$) layers. Its effective Hamiltonian $H$ is distinct from the Hamiltonian $H_0$ of the isolated layer, acquiring properties from $t$ and $b$. **b**, Device geometry to investigate proximity SOC in bilayer graphene (BLG) encapsulated with $WSe_2$. The charge density $n$ and displacement field $D$ are controlled with top and bottom gates. **c**, Penetration field capacitance $C_p$ as a function of $n$ and $D$ showing an incompressibility peak at $D = 0$ for $n = 0$. Arrows indicate maxima associated to band-splitting due to proximity-SOC. **d**, Illustration of the Dzyaloshinskii-Moriya interaction (DMI) at the vdW interface of a 2D magnet ($Fe_3GeTe_2$) and a TMDC with large SOC ($WTe_2$), $s_i$ and $s_j$ represent the spins of neighboring atoms and $d_{ij}$ the DMI vector. The interfacial DMI stabilizes magnetic skyrmions, as observed by Lorentz transmission electron microscopy in $WTe_2$-$Fe_3GeTe_2$ (magnetic field: 0.51 T; scale bar: 500 nm). **e**, Schematics of a lateral spin device for investigating proximity SOC in graphene by means of spin transport. The device consists of a graphene spin channel, which is partially covered with a TMDC. The two attached ferromagnets (F1 and F2) act as spin injector and detector, respectively. A charge current $I$ through F1 injects spins with an orientation parallel to the F1 magnetization direction. While diffusing towards the spin detector (F2), the spins precess under either an oblique or in-plane magnetic field $B$, as represented in the top and bottom insets. When they reach the TMDC, their orientation is characterized by the angle $\beta^*$, whose value is with the magnitude of $B$. **f**, Spin precession experiments in a graphene-$WS_2$ device. The nonlocal signal $R_{NL}$ is shown for $B$ perpendicular to the plane (blue) and in-plane (red) for which spins stay in the graphene plane and precess out of it, respectively (see insets). The disparity of the curves demonstrates the highly anisotropic nature of the spin dynamics in graphene-$WS_2$. Panels **b**-**c** and **e**-**f** are adapted from Ref.[88] and Ref.[37], respectively. SNL. The Lorentz microscopy image in **d** is from Ref.[89], SNL.

Graphene proximitized by a TMDC can also exhibit an inverted band structure (Box 2 b), suggesting emerging topological phenomena[70,73,90], distinct from isolated graphene[91] and driven by the valley-Zeeman coupling. Although the band structure remains topologically trivial, protected pseudo-helical states appear at zigzag edges of proximitized nanoribbons. An inverted band structure in BLG-WSe$_2$ stacks has been confirmed experimentally[88] (Fig. 2b and 2c). Topological quantum spin Hall phases were also predicted in proximitized BLG[92], while helical edge modes in BLG-WSe$_2$ heterostructures were recently reported[93].

Proximity SOC is becoming increasingly important in 2D materials beyond graphene. A SOC enhancement could help stabilize the anisotropy or the magnetic order of a 2D magnet. An increase in the Curie temperature $T_C$ of Fe$_3$GeTe$_2$ (FGT) up to 400 K has been observed when grown onto Bi$_2$Te$_3$. The larger $T_C$ in thinner FGT films on Bi$_2$Te$_3$, when the opposite trend is observed in FGT alone, suggests the presence of an interfacial effect[94], although it is unclear why a significant $T_C$ increase is observed in relatively thick FGT films (up to 10s of nm). Similarly, an increase in $T_C$ was observed[95] in the Heisenberg ferromagnet V$_5$Se$_8$ when in contact to NbSe$_2$. The enhanced $T_C$ was accompanied by a strong out-of-plane magnetic anisotropy and was attributed to the Zeeman SOC in NbSe$_2$. A large SOC together with broken inversion symmetry can also favour the anti-symmetric Dzyaloshinskii–Moriya exchange interaction (DMI) and lead to topological magnetic configurations[96] (Fig. 2d). Néel-type skyrmions were observed in WTe$_2$-FGT using Lorentz transmission electron microscopy[89] (Fig. 2d); the large interfacial DMI energy of ~1.0 mJm$^{-2}$ was attributed to induced Rashba SOC.

**Anisotropic spin relaxation and spin filtering.** One of the first spin-device realizations combining graphene with a TMDC comprised a graphene lateral spin device partially capped with MoS$_2$. Using electrostatic gating, the spin current across the graphene channel was controlled between ON and OFF states, a phenomenon attributed to spin absorption at the MoS$_2$[35]. It is argued that in the OFF-state spins could move freely between graphene and MoS$_2$, due to the gate-induced suppression of the Schottky barrier between graphene and $MoS_2$, leading to fast spin relaxation[35,97].

Further studies[36,37] reported anisotropic spin relaxation in graphene-TMDC heterostructures (with TMDC = $MoSe_2$, $MoS_2$ and $WS_2$), even in the absence of spin absorption[37]. By implementing out-of-plane spin precession techniques[59–61] (Figs. 2e and 2f) the spin relaxation anisotropy ratio $\zeta = \frac{\tau_{s,\perp}}{\tau_{s,\parallel}}$ were quantified. It was observed that the in-plane spin component is strongly reduced when propagating through the graphene-TMDC region, with $\tau_{s,\parallel}$ in the range of a few picoseconds (two orders of magnitude smaller than in reference graphene devices[37]). In contrast, the out-of-plane spin component propagates much more efficiently, with $\tau_{s,\perp}$ in the range of tens of picoseconds[36,37] and thus $\zeta \sim 10$. These results evidence that graphene-TMDC heterostructures act as spin filters, whose spin transmission is tailored by the spin orientation.

When no spin current is absorbed by the TMDC, the anisotropy can be fully attributed to proximity-induced SOC[37,60]. According to theoretical predictions, the spin dynamics is controlled by the spin-valley coupling imprinted onto graphene[98,99]. The spin relaxation is governed by the Dyakonov-Perel mechanism, with $\tau_{s,\perp}$ and $\tau_{s,\parallel}$ largely determined by the momentum ($\tau_p$) and intervalley ($\tau_{iv}$) scattering times, respectively, typically with $\tau_p \ll \tau_{iv}$. Because of the relatively long $\tau_{iv}$, the in-plane spins precess under a slowly fluctuating effective (perpendicular) magnetic field between $K$ and $K'$, leading to fast spin relaxation. In contrast, because of the short $\tau_p$, out-of-plane spins precess under fast fluctuating Rashba fields and their relaxation is suppressed due to motional narrowing[13]. The spin-anisotropy ratio, derived from the emergent Hamiltonian (Box 2), is[98–100]

$$\zeta = \left(\frac{\lambda_{VZ}}{ak\Delta_{PIA} \pm \lambda_R}\right)^2 \frac{\tau_{iv}}{\tau_p} + \frac{1}{2} \approx \left(\frac{\lambda_{VZ}}{\lambda_R}\right)^2 \frac{\tau_{iv}}{\tau_p} + \frac{1}{2} \qquad (1)$$

with the approximation being valid about the Dirac point or for small $\Delta_{PIA}$ (Box 2). In the absence of valley-Zeeman SOC, $\zeta = \frac{1}{2}$ with out-of-plane spins relaxing faster than in-plane spins, as expected for a 2D Rashba system[56]. Using an interband tunnelling description and first-principles calculations, it has been proposed that the SOC strength can be tuned with the Fermi energy, resulting in an energy-dependent anisotropy[77,78].

Anisotropic spin relaxation has also been discussed theoretically in graphene-TIs[81] and graphene-hBN heterostructures[74]. Moreover, $\zeta \sim 10$ has been measured in hBN-BLG-hBN at temperatures around 100 K near the charge neutrality point[101,102]. The spin relaxation becomes isotropic either at large enough carrier densities or at high temperatures (at room temperature $\zeta \sim 1$). Similar to graphene-TMDC, the large $\zeta$ seems to arise from the spin-valley coupling associated to the intrinsic SOC in BLG.

**Charge-to-spin interconversion.** CSI phenomena driven by SOC are amongst the most relevant effects in modern spintronics[103]. Their presence can reveal subtle spin-orbit interactions and spin dynamics in the investigated materials. They are also central for next generation SOT magnetic memories (SOT-MRAM)[9,10] as well as for proposals targeting energy-efficient spin-logic architectures[104] (Box 1). CSI in 2D materials has been gaining increasing attention, following the report of SOTs with non-trivial (and potentially useful) symmetries using TMDCs[105], the achievement of magnetization switching with TIs and TMDCs[106,107], as well as the observation of spin Hall effects (SHE) and spin galvanic effects (SGE) [108–111] (Fig. 3).

A recent surge of experiments on vdW heterostructures has been triggered by the use of graphene as a channel to transport a spin current from a ferromagnetic contact to the CSI region[112,113]. The device geometry is analogous to that developed for fully metallic systems[103], consisting of a graphene Hall cross with a large SOC 2D material along one of the arms and ferromagnetic injector and detector contacts across the other (Fig. 3a). The first experiments using platinum (Pt), a well-known material with efficient CSI by the SHE[103], demonstrated large

CSI[112,113] and established the use of spin-precession to investigate the nature of the CSI in 2D heterostructures[113]. The analysis and interpretation of the results differ for insulating or conducting SOC materials. While in the former case it is possible to directly ascribe the CSI to proximity-induced SOC, in the latter case the overall CSI can aggregate the CSI arising from proximity effects and the CSI at the surface and/or bulk of the conducting SOC material. The anomalously large CSI in graphene-Pt could be due to such aggregation of effects, although this remains to be clarified[113].

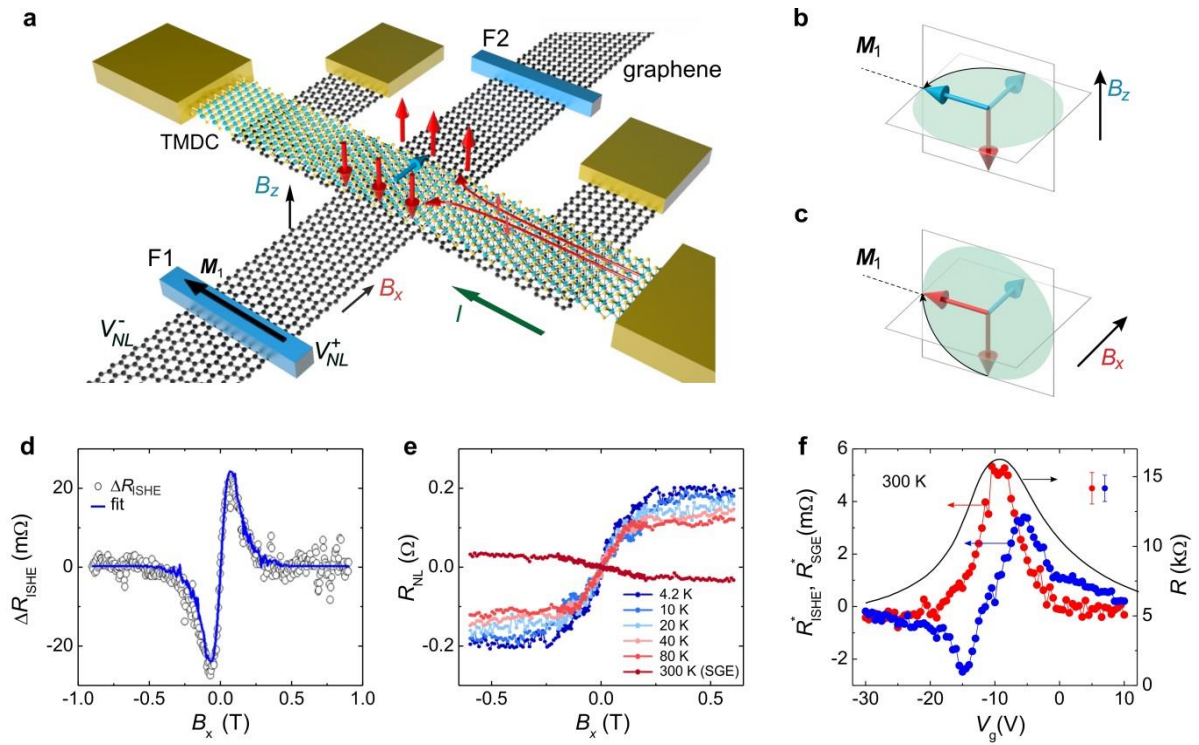

**Fig. 3 | Charge-to-spin interconversion (CSI) in vdW heterostructures. a,** Schematics of a lateral spin device for CSI experiments. The device consists of a graphene Hall cross with a TMDC covering one of the arms, which enhances the SOC in graphene. A current $I$ along the graphene arm creates a transversal spin current with out-of-plane spin polarization driven by the spin Hall effect (SHE). Simultaneously, a non-equilibrium spin density, with in-plane spin polarization, is built at the proximitized graphene driven by the inverse spin galvanic effect (ISGE). Spin-sensitive contacts F1 and F2 probe diffusing spin currents under an applied magnetic field $B$. For $B$ applied along $z$ ($B_z$) and along $x$ ($B_x$) only the spins arising from the ISGE (**b**) and the SHE (**c**) precess, respectively. **d,** Room temperature ISHE signal $\Delta R_{ISHE}$ as a function of magnetic field $B_x$ in a device with MoS$_2$ on few-layer graphene. **e**, SGE signal measured at different temperatures in a graphene-WS$_2$ device. **f**, Experimental demonstration of the coexistence and gate tunability of the ISHE and the SGE in graphene-WS$_2$ at room temperature. Panels **a-c** and **f** adapted from Ref.[110] SNL. Panels **d** and **e** adapted from Ref.[109] and Ref.[111], respectively, American Chemical Society.

CSI driven by SOC in a vdW heterostructure was first confirmed in multilayer-graphene-MoS$_2$[109] (Fig. 3d). This report was soon followed by the simultaneous observations of the SHE and SGE in graphene–WS$_2$[110,111] (Figs. 3e and f). The CSI in graphene by proximity SOC[114–116] is best established by ruling out the spin absorption in the TMDC[35,97,110]. Notably, the CSI can be controlled upon electrostatic gating, which tunes the graphene carrier density *n*. A gate dependent CSI in proximitized graphene was first observed with WS$_2$ up to 75 K for the inverse SGE[111] and up to room temperature for the inverse SGE and SHE (and reciprocal effects)[110]. Gate dependence of the inverse SHE and of the SGE was later on reported in graphene-WSe$_2$, and in graphene-TaS$_2$ and graphene-(Bi,Sb)$_2$Te$_3$, respectively[117–119].

The effective conversion efficiencies compare favourably with those of metallic systems[109,110,118]. Furthermore, the experimental dependence of the CSI vs *n* in graphene-WS$_2$[110] agrees with theoretical modelling, for both proximity-induced SGE[115] and SHE[114]. The SOC strength has been estimated using the Kubo-Bastin formula[110,114]. By matching model calculations with the experimental results $\lambda_\mathrm{I} \sim 0.2$ meV and $\lambda_\mathrm{vZ} \sim 2.2$ meV are obtained[110] (Box 2). Prior reports of SHE in graphene by proximity of WS$_2$ suggested a much larger SOC (17 meV)[120]. However, these experiments used the so-called H-geometry[103], which in graphene devices is sensitive to a variety of phenomena, not necessarily spin related[5,121–123].

CSI was also investigated in other conducting 2D materials, following the same approach as with Pt[112,113]. Experiments using 1T'-MoTe$_2$ revealed an unconventional CSI in which a charge current arises parallel to the spin orientation[124]. It is unclear whether the CSI originates in the bulk or the surface of the material. The observation is reminiscent to the appearance of unconventional SOTs in low-symmetry WTe$_2$[105,125], suggesting that the crystalline mirror symmetry of 1T'-MoTe$_2$ is broken, perhaps by strain introduced during device fabrication. Unconventional CSI was also observed in WTe$_2$, with an efficiency approaching 10 %; control experiments indicate that the CSI originate in the bulk of the material[126].

**Magnetic proximity effects**

When non-magnetic 2D materials, such as graphene or TMDCs, are in contact with a magnetic material, they can experience a proximity-induced MEI. The induced magnetism is characterized by a net local spin polarization in equilibrium and an energy splitting of the bands, which in graphene is equal at different valleys (in the absence of SOC). The proximity MEI is parametrized by the exchange coupling strength $\lambda_{EX}$ when the non-magnetic 2D material is in contact with a ferromagnet and $\lambda_{EX}^{AF}$ when in contact with an antiferromagnet (Box 2). Typically, the goal is to achieve a large $\lambda_{EX}$ ($\lambda_{EX}^{AF}$) while maintaining the (spin) transport capabilities of the isolated layer.

**Proximity MEI in graphene.** Early first-principles calculations predicted a $\lambda_{EX}$ of tens of meVs in graphene when proximitized by bulk materials such as EuO[127] or the ferrimagnet $Y_3Fe_5O_{12}$ (YIG)[128]. Similar $\lambda_{EX}$ were estimated with conventional ferromagnetic metals, such as Co or Fe, across a thin hBN insulating barrier[8,129,130]. Control of proximity exchange by electrical polarization has been predicted in graphene on multiferroic $BiFeO_3$[131]. The first experimental results were also reported in graphene proximitized by bulk materials, albeit with typical $\lambda_{EX}$ values that were significantly smaller than expected. Charge transport experiments in graphene-YIG showed the presence of an anomalous Hall resistance[132], while Zeeman SHE indicated an exchange field of up to 14 Tesla (1.5 meV) in graphene-EuS[133]. Subsequently, spin transport experiments, using lateral devices based on graphene[134] and BLG[135] onto YIG, provided more direct indications of proximity MEI and demonstrated spin current modulation (Fig. 4a and 4b), although $\lambda_{EX}$ was found to be even smaller[134] (~20 μeV). Proximity MEI was also reported in YIG-graphene-hBN through nonlocal charge transport measurements[136], Co-graphene-NiFe junctions[137], and gate-dependent spin-inversion in edge-contacted graphene spin valves[23].

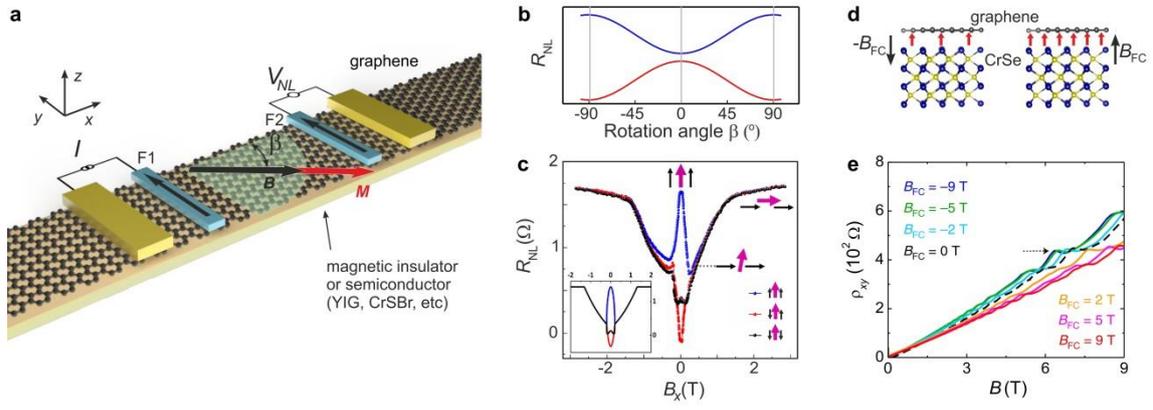

**Fig. 4 | Magnetic proximity effects in graphene. a,** Schematics of experimental device to detect exchange interaction in graphene, induced by proximity of a magnetic substrate using spin transport. A magnetic field ***B*** sets the orientation of the substrate magnetization ***M***, and thus of the proximity-induced exchange field ***B***$_{EX}$ in graphene. **b,** Representation of $R_{NL} = \frac{V_{NL}}{I}$ versus β in parallel (blue) and antiparallel (red) alignment of the magnetizations of the ferromagnetic contacts F1 and F2. When β= 0, ***B***$_{EX}$ is perpendicular to the orientation of the injected spins, and the signal presents the minimum magnitude. The coercive field of the substrate is assumed to be much lower than that of F1 and F2, as in the case of YIG. Therefore ***M*** follows ***B*** while the magnetizations of the contacts remain along their long axis (black arrows). **c,** $R_{NL}$ as a function of magnetic field along *x* (see **a**) in a graphene-CrSBr heterostructure. The initial arrangement of the magnetizations of the injector and detector contacts (black arrows) and the substrate (purple arrow) is achieved with ***B*** along *y*. In this case, the coercive filed of the contacts is smaller than that of the substrate (an antiferromagnet). The inset shows calculated precession curves with a spin-dependent graphene conductivity model. **d,** Schematics of graphene coupled to CrSe. The red arrows represent the remnant magnetization induced at the interface by field cooling with opposite magnetic field orientations. **e,** Exchange splitting in graphene on top of antiferromagnetic CrSe. Quantum Hall plateaus are shifted by field cooling in the heterostructure. The black arrow denotes the most pronounced plateau shifted by field cooling. The shift in the plateau is attributed to the change in the remnant magnetization. Panel **c** adapted from Ref.[138]. Panels **e** adapted from Ref.[139], SNL.

The small proximity $\lambda_{EX}$ observed with bulk magnets could be ascribed to rough interfaces, thus recent Investigations have shifted towards proximity MEI by 2D magnets, which promise atomically smooth interfaces. Relevant 2D ferromagnets include the $Cr_2X_2Te_6$ (X= Si, Ge or Sn) or $CrX_3$ (X=I, Br or Cl) families with predicted $\lambda_{EX}$ in the range of several meVs[72] (Box 2 d). The induced MEI with antiferromagnets, such as $MnPSe_3$, a 2D Heisenberg-like antiferromagnet, could lead to sub-meV staggered exchange coupling in graphene[73] (Box 2 e). A few experiments indeed indicate a significant proximity MEI[138–140]. In graphene-CrSBr, with CrSBr an interlayer antiferromagnet, charge and spin transport driven by electrical bias and thermal gradients[138] indicate $\lambda_{EX}$ ~ 20 meV, corresponding to an exchange field of ~ 170 T at 4.5 K (Fig. 4c). Proximity MEI was also reported in graphene-CrSe heterostructures[139], where CrSe is a non-collinear antiferromagnet with a complex phase diagram. A magnetized interface (Fig. 4d), which does not occur in CrSe alone, is observed in graphene-CrSe with transport and magneto-optic measurements after magnetic-field cooling. The proximity exchange field was quantified from shifts in the quantum Hall plateaus and quantum oscillations (Fig. 4e), resulting in $\lambda_{EX}$ larger than 130 meV at 2 K.

**Proximity MEI beyond graphene.** Proximity MEI has also been investigated in materials such as TMDCs and TIs. Experiments in semiconducting TMDCs typically rely on optical techniques. In $WSe_2$-EuS[141] and $WS_2$-EuS[142] heterostructures, the reflection and photoluminescence spectra of circularly-polarized photons probe the electronic states of the TMDC and quantify the proximity-induced exchange splitting. For $WSe_2$, it was estimated that $\lambda_{EX}$ ~ 2-4 meV, corresponding to an exchange field of ~10-20 T. For $WS_2$, the splitting was found to be much larger $\lambda_{EX}$ ~ 19 meV and to have opposite sign. According to theoretical modelling[142], the magnitude and sign of the splitting is determined by the surface termination of EuS and the band alignment between TMDCs and EuS.

As with graphene, a growing number of studies are being carried out with 2D magnets[143,144]. Placing a monolayer TMDC on $CrI_3$ (Fig. 5a) results in an estimated exchange splitting in the meV range[145], which should affect the exciton spectra[69]. Given its short range interaction, the proximity effect allows probing the magnetization of the 2D magnet adjacent layer, even in the absence of a global magnetic moment. A layer-dependent magnetic proximity effect has been observed in monolayer $WSe_2$ on few-layer $CrI_3$[144]. While magneto-optic measurements demonstrate that bilayer $CrI_3$ is a layered antiferromagnet, circularly-polarized photoluminescence shows that the exchange splitting in $WSe_2$ is most sensitive to the interfacial layer. The contribution of the second layer to the splitting is of substantially smaller magnitude and has an unexpected opposite sign (Figs. 5b and 5c). The quantitative interpretation of the exciton spectra and dynamics is not straightforward. The hybridization of the TMDC orbitals with the spin-polarized $CrI_3$ orbitals is rather complex. In particular, it is expected that twisting the two layers would lead to significant variations of the proximity MEI, both in magnitude and character[145]. Furthermore, photoluminescence studies in $MoSe_2$-$CrBr_3$ uncovered a charge dependence of the proximity effects in which the valley polarization of the $MoSe_2$ trion state follows the local $CrBr_3$ magnetization, while the neutral exciton state is insensitive to it[146]. This is attributed to spin-dependent interlayer charge transfer in timescales between the exciton and trion radiative lifetimes.

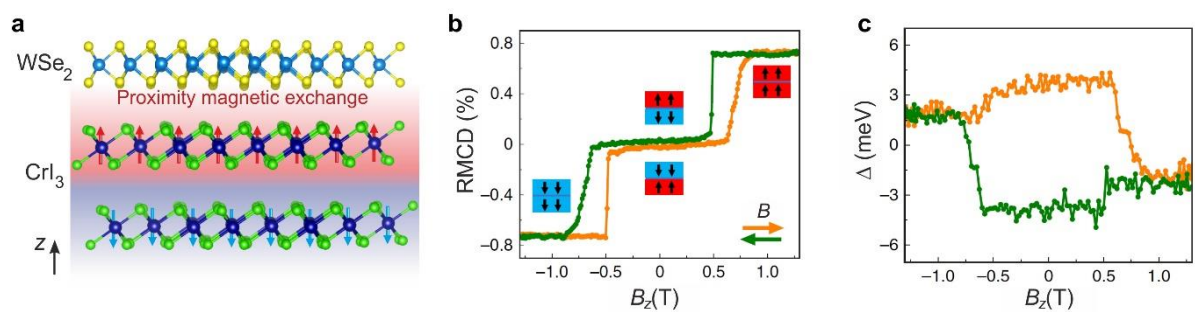

**Fig. 5 | Magnetic proximity effects in TMDCs. a,** Schematic of monolayer $WSe_2$ interfacing with a magnetic layer ($CrI_3$ in this case) to investigate proximity control of valley dynamics. **b,** Reflexion magnetic circular dichroism (RMCD) as a function of magnetic field in $WSe_2$-$CrI_3$, showing typical features of a layered antiferromagnetic bilayer $CrI_3$. **c,** Valley Zeeman splitting $\Delta$ as a function of magnetic field as determined from photoluminescence measurements. Panels **b**-**c**, are adapted from Ref.[143], SNL.

Magnetic proximity effects are also intensively investigated in layered 2D and 3D TIs. In the 3D TIs, such as the $Bi_2Te_3$ family, broken time-reversal symmetry induces a gap in the Dirac band dispersion of the surface states[147]. Tuning of the Fermi level in the gap leads to the emergence of a quantum Hall effect at zero magnetic field: the quantum anomalous Hall effect (QAHE), a phenomenon that is very promising for quantum metrology. Signatures of proximity magnetism in 3D TIs have been reported, for example, in [EuS, YIG, $Tm_3Fe_5O_{12}$, $Cr_2Ge_2Te_6$]/TI with the observation of an anomalous Hall effect[148–150] or by investigating spin-polarized neutron reflectivity[151]. However, the origin of the magnetic signals in these type of experiments is usually not fully understood[152–154]. An unambiguous demonstration of proximity MEI was reported (Zn,Cr)Te/(Bi,Sb)$_2$Te$_3$/(Zn,Cr)Te heterostructures with the observation of the QAHE[155]. In 2D TIs, or quantum spin Hall insulators, such as monolayer $WTe_2$, the conduction is dominated by helical edge states[156,157] with canted spin orientation due to reduced symmetries[158–160]. When WTe$_2$ is placed in a heterostructure with CrI$_3$, magnetic proximity could lead to a change in the edge state conductance that is controlled by the magnetization of the interfacial CrI$_3$ layer[161]. This could ultimately result in the observation of the QAHE, depending on other phenomena, such as charge transfer at the interface.

**Conclusions and future perspectives**

Recent progress in the design of complex vdW heterostructures brings unprecedented possibilities for developing innovative ultra-compact spin devices and computing architectures. In respect to conventional spintronic applications, it is necessary to identify the best combination of 2D materials to demonstrate practical magnetic tunnel junctions (either with conducting or insulating 2D magnets) or CSI-induced switching of 2D magnets[162] (using high spin-orbit materials such as WTe$_2$ and Bi$_2$Te$_3$). Voltage control of magnetic properties is another promising avenue available with 2D materials (Box 1). Advances have been made in this regard[7,38], while electric-field dependent proximity SOC in 2D magnets will certainly bring further opportunities. In addition, the reduced symmetries in monolayer TMDCs such as MoTe$_2$ and WTe$_2$ lead to persistent spin textures, which result in multicomponent SHE[124,126]. In combination with the externally tunable spin-orbit fields, they may enable electric control of SOTs[163].

Proximity effects can be further exploited for novel spin-orbit[164,165] or magnetic valves[166,167] comprising BLG and a large SOC material or a 2D magnet, respectively. The interplay of two factors -- short-range of the proximity effect and the layer polarization in BLG -- results in layer-polarized electronic bands and asymmetrically conduction and valence bands. Applying a transverse electric field can reverse this situation, turning the SOC (or $\lambda_{EX}$) ON or OFF by electric field or doping and leading to novel spintronic functionalities[13]. An additional functionality is offered by intercalating BLG between 2D magnets, forming a spin valve that could resolve parallel and antiparallel magnetizations in transport[168]. Another exciting perspective is engineering both the SOC and MEI, and their interplay, as in the so called ex-so-tic vdW heterostructures (Box 2 f). In graphene, such an interplay is predicted to induce the QAHE[169], novel topological phases[73,170], proximity-based SOT[72], unique signatures of anisotropic magnetoresistance, and even new functionalities based on swapping spin-orbit coupling and exchange, all in a single device[171]. The experimental observation of these phenomena will be key milestones in spintronics and quantum metrology.

The control of interlayer twist between layers can be further exploited to tailor the spin interactions (Box 1). For instance, the atomic stacking in the Moiré pattern in twisted $CrBr_3$ bilayers modulates the proximity SOC and MEI, as revealed by spin-polarized scanning tunnelling microscopy[172]. This and other emerging phenomena could become mainstream in the forthcoming years for 2D spintronics, including the control of information transfer via magnons in 2D magnets[173,174] and topological magnetic structures like skyrmions and Néel spin spirals, which have been already observed in FGT and FGT-based heterostructures[89,175–177] and predicted in 2D Janus materials[178].

Many of these technological prospects will require overcoming important challenges. A particularly critical one is the development of large-area, stable 2D magnets with magnetic order at room temperature, using scalable stacking and growth processes. Proximity SOC has shown potential to increase $T_C$; similarly, a significant enhancement of $T_C$ could also be achieved by coupling a 2D ferromagnets to an antiferromagnet, as in $Fe_3GeTe_2$-$FePS_3$ heterostructures[179]. Taking advantage of SOTs will demand a full understanding of the mechanisms for (vertical) spin transfer across heterostructures and ways to take advantage of the SOTs induced by 2D materials with reduced symmetries. Applications relying on topological phases, such as the QAHE, also require robust magnetic properties to both applied currents and high temperatures. In regards to skyrmions, many fundamental challenges lie ahead beyond material issues, such as the development of writing, processing and reading functionalities using all-electrical schemes. Moreover, increasingly realistic theoretical modelling of proximity effects in complex vdW heterostructures (combining different 2D material families) is necessary to grasp the subtle spin and exciton dynamics and to separate the contributions of the exchange interaction from spin and orbital moments. Extracting (minimum) model Hamiltonian from *ab-initio* calculations is becoming very challenging owing to the intertwined combination of all interactions involved, necessary for performing spin transport simulations. Precise comparison with experiments is hampered by the difficulty of reproducing interfaces and controlling the stacking, especially in multilayer heterostructures.

The advances covered in this Review therefore represent the starting point for 2D material design for spintronics and optospintronics. However, the endless possibilities offered by proximity effects promise an enduring impact in terms of innovative devices and architectures. Engineering vdW heterostructures can reveal novel classes of artificial quantum materials[180], offering opportunities for both scientific discoveries and technological breakthroughs. Information and quantum computing paradigms, electrically driven light emitters, photodetectors, and sensors might emerge by harnessing the rich internal degrees of freedom of 2D materials (spin, valley, sublattice, excitonic and layer pseudospin) and the creation and manipulation of entangled states.

**Acknowledgements**

We thank D. Torres (ICN2) for help in implementing the 3D device models used in the figures. J.F.S. and S.O.V. acknowledge support of the European Union's Horizon 2020 FET-PROACTIVE project TOCHA under grant agreement 824140, the King Abdullah University of Science and Technology (KAUST) through award OSR-2018-CRG7-3717, and MINECO under contracts no. PID2019-111773RB-I00/AEI/10.13039/501100011033 and no. SEV-2017-0706 Severo Ochoa. J.F., S.R. and S.O.V. acknowledge support of the European Union Horizon 2020 research and innovation program under contract 881603 (Graphene Flagship) and J.F. of the Deutsche Forschungsgemeinschaft (DFG, German research Foundation) SFB 1277 (project-id:314695032) and SPP 2244. R.K.K. acknowledges support of US DOE-BES (DE-SC0016379), AFOSR MURI 2D MAGIC (FA9550-19-1-0390), and NSF MRSEC (DMR-2011876).

**Competing Interests**

The authors declare no competing interests

**Additional Information**

Correspondence should be addressed to J.F.S. or S.O.V.